\documentclass[12pt]{article}
 \pdfoutput=1
 \usepackage{graphicx}
  \usepackage{epsfig}
  \usepackage{amsmath}
  \usepackage{amssymb}
  \usepackage{color}
 \usepackage{dcolumn}
\usepackage{amsbsy}
\usepackage{bm}
\usepackage{amsthm}
\usepackage[caption=false]{subfig}
\usepackage{tabularx}
\usepackage{array}
\usepackage{hyperref}
  \newlength{\abstractwidth}
  \renewcommand{\title}[1]{\vbox{\center\bf{\Large{#1}}}\vspace{5mm}}
  \renewcommand{\author}[1]{\vbox{\center#1}\vspace{5mm}}
  \newcommand{\address}[1]{\vbox{\center\em#1}}
  \newcommand{\email}[1]{\vbox{\center\tt#1}\vspace{5mm}}

\begin{document}

\begin{titlepage}
\begin{center}
\hfill \\
\hfill \\
\vskip 1cm

\title{``Almost" Quotient Space, Non-dynamical Decoherence and Quantum Measurement}

\author{Yu-Lei Feng$^{1}$ and Yi-Xin Chen$^1$}

\address{$^1$Zhejiang Institute of Modern Physics, Zhejiang University, Hangzhou 310027, China}

\email{11336008@zju.edu.cn} \email{yxchen@zimp.zju.edu.cn}

\end{center}

  \begin{abstract}
 An alternative approach to decoherence, named non-dynamical decoherence is developed and used to resolve the quantum measurement problem. According to decoherence, the observed system is open to a macroscopic apparatus(together with a possible added environment) in a quantum measurement process. We show that this open system can be well described by an ``almost" quotient Hilbert space formed phenomenally according to some stability conditions. A group of random phase unitary operators is introduced further to obtain an exact quotient space for the observed system. In this quotient space, a density matrix can be constructed to give the Born's probability rule, realizing a (non-dynamical) decoherence. The concept of the (``almost") quotient space can also be used to explain the classical properties of a macroscopic system. We show further that the definite outcomes in a quantum measurement are mainly caused by the ``almost" quotient space of the macroscopic apparatus.

  \end{abstract}

\end{titlepage}

\tableofcontents

\baselineskip=17.63pt








\section{Introduction}
\label{seci}
Quantum measurement is an important problem to understand quantum mechanics, in particular it is significant to reconcile the microscopic quantum world with the macroscopic classical world perceived by us. The key distinction is the superposition principle in quantum mechanics, which says that any linear combination of a system's states are still a possible state of that system. But this principle seems not to be suitable for the ordinary macroscopic systems perceived by us. For example in a measurement, the outcomes that the apparatus records must be definite, not a superposition of them, indicating that the state of the observed system will also become definite after the measurement, i.e. \emph{the problem of definite outcomes}. In the familiar Copenhagen Interpretation, this is explained by a so called ``state collapse" of the observed system, a mysterious explanation\footnote{In the Copenhagen Interpretation, the ``state collapse" is regarded to be a physical process that occurs instantaneously, which violates the principle of locality. In the view of dynamical decoherence, there is a decoherence time so that the ``state collapse" occurs not instantaneously. In our approach, however, the ``state collapse" is not a physical process, but only a coarse grained description.}.

Another explanation is provided by quantum decoherence~\cite{a,b}. The key idea is that realistic quantum systems are never isolated, but are immersed in the surrounding environment and interact continuously with it. Then, ``decoherence is caused by the interaction with the environment which in effect monitors certain observables of the system, destroying coherence between the pointer states corresponding to their eigenvalues"~\cite{a}. An \emph{environment-induced superselection or einselection} scheme~\cite{c} has been developed to handle the measurement problem\footnote{There are also some other attempts to the quantum measurement problem. Since our approach is according to the concept of decoherence, we restrict our discussions only within this attempt, in which einselection is one particular representative scheme.}, saying that the apparatus's pointer states are determined or selected by the form of the interaction between the apparatus and its environment.

The einselection is briefly as follows. Suppose the observed system's initial state is $\alpha\vert S_1\rangle+\beta\vert S_2\rangle$. After a measurement, we obtain a correlated state for the combined system-apparatus
\begin{equation}
\alpha\vert S_1\rangle\vert A_1\rangle+\beta\vert S_2\rangle\vert A_2\rangle.
\label{a}
\end{equation}
There is a basis ambiguity that we can rewrite the above correlated state in any other basis, the so called \emph{preferred-basis problem}. For example, we could combine the apparatus's states $\{\vert A_1\rangle,\vert A_2\rangle\}$ linearly to obtain another couple $\{\vert A'_1\rangle,\vert A'_2\rangle\}$ so that the corresponding states for the system $\{\vert S'_1\rangle,\vert S'_2\rangle\}$ may not be the eigenstates of the measured observable. This is possible in principle due to the superposition principle. The einselection scheme tries to resolve this problem by coupling the apparatus with an environment, obtaining a more correlated state
\begin{equation}
\alpha\vert S_1\rangle\vert A_1\rangle\vert E_1\rangle+\beta\vert S_2\rangle\vert A_2\rangle\vert E_2\rangle.
\label{b}
\end{equation}
Moreover, a \emph{stability criterion} $[\hat{O}_A,\hat{H}_{A,E}]=0$ was suggested by Zurek~\cite{c} so that the system-apparatus correlations are left undisturbed by the subsequent formation of correlations with the environment, i.e. there is a preferred apparatus observable. Then provided $\langle E_1\vert E_2\rangle\simeq0$, a reduced density matrix for the system-apparatus can be obtained by performing a partial trace over the environment's space
\begin{equation}
|\alpha|^2\vert S_1\rangle\langle S_1\vert\vert A_1\rangle\langle A_1\vert+|\beta|^2\vert S_2\rangle\langle S_2\vert\vert A_2\rangle\langle A_2\vert,
\label{c}
\end{equation}
which is a statistical ensemble with definite outcomes, realizing the Born probability rule for the observed system.
Although the einselection scheme gives the required results in Eq.~(\ref{c}), there are still some problems with this scheme, for example the method to obtain the Born rule seems to be a \emph{circular argument}~\cite{a,b}. A more detailed discussion including a comparison between our approach and the einselection scheme, will be given in Sec.~\ref{secvii}.

For a large closed system, suppose one small portion or subsystem is singled out to be studied or observed, then the rest subsystems can serve as an environment. More precisely, these can be expressed as
\begin{equation}
\mathcal{H}_t=\mathcal{H}_S\otimes\mathcal{H}_E,
\label{d}
\end{equation}
with $\mathcal{H}_t$, $\mathcal{H}_S$ and $\mathcal{H}_E$ the Hilbert spaces of the total closed system, the studied system and the environment respectively. The evolution of the studied system can be described by a superoperator~\cite{d}, obtained from some total evolution $\hat{U}_t$ by tracing over the environment's space. This gives a coarse grained description to the studied system. Moreover, the quantum information stored in the correlations with the environment may be ``lost" in the following way. The full reduced density matrix in Eq.~(\ref{c}) in fact has a factor $\langle E_i(t)\vert E_j(t)\rangle$, where the time denotes an explicit dynamical evolution. When $i=j$, this factor can be assumed to be one according to normalization of the states. While the remaining factor $\langle E_1(t)\vert E_2(t)\rangle$ may be related to a so called \emph{decoherence function} $\Gamma_{12}(t)$~\cite{d1}
\begin{equation}
|\langle E_1(t)\vert E_2(t)\rangle|^2=\exp\Gamma_{12}(t),\qquad \Gamma_{12}(t)\leq 0.
\label{e}
\end{equation}
This decrease or decay leads to the orthogonality condition $\langle E_1\vert E_2\rangle\simeq0$, realizing the statistical ensemble in Eq.~(\ref{c}). This kind of \emph{dynamical decoherence} is found to occur in some controlled models~\cite{d1}, where the environment is usually assigned to be thermal or possess some randomness. In this sense, the environment seems to be more classical, violating the Everettian picture~\cite{d2} in which only quantum unitary dynamics occur in a closed system~\cite{d3}. This can also be seen directly via the decay term in Eq.~(\ref{e}) that gives some kind of non-unitary dynamical evolutions. The non-unitary dynamical evolution can also occur in the mast equation~\cite{d}, which is derived through a Markovian approximation that ignores the feedback from the studied system to its environment. This means that this kind of approximation is not consistent with the Everettian picture. In Sec.~\ref{secii}, we will introduce a random phase approximation that can be described in a way consistent with the Everettian picture.

The information is actually not lost from the view of the total closed system within the Everettian picture. To demonstrate this fact, we'd better to always work on the total Hilbert space $\mathcal{H}_t$. Then how to obtain a coarse grained description of the studied system \emph{without partially tracing} over the environment's degrees of freedom? Recall that the total Hilbert space is closed under all the unitary transformations, which can be roughly divided into two classes: one class contains non-dynamical representation transformations that are similar to coordinate transformations; while the other one contains all the dynamical unitary evolutions. Those representation transformations can be neglected by choosing a particular representation or gauge. After that, we can just consider the dynamical evolutions. For a closed system, its relevant space can be \emph{ergodic} from one particular state under all the evolutions that are commutative with the Hamiltonian. In this way, the total Hilbert space is \emph{coarse grained} into some constant energy hypersurfaces under the ``detection" of those evolutions. That is, if we apply \emph{a smaller subset of evolutions satisfying some given condition}, the fine structure of the total Hilbert space may not be ``detected" so that only a \emph{coarse grained} description can be obtained. In this paper, we shall show that both a quantum measurement process and a macroscopic system can be described in this way, without using a partial trace.

The paper is organized as follows. In Sec.~\ref{secii1}, an ``almost" quotient space is formed to describe a particular open system, an observed system open to a measurement apparatus. This ``almost" quotient space and a following random phase approximation given in Sec.~\ref{secii2} provide a coarse grained description to the observed system, realizing a non-dynamical decoherence. Then in Sec.~\ref{seciii}, a group of random phase unitary operator is introduced to obtain an effective quotient space to describe the observed system. In that quotient space, the superposition principle is not proper, and a density matrix for a statistical ensemble is obtained to give the Born rule. A two-dimensional example, the Stern-Gerlach experiment is analyzed in Sec.~\ref{seciv} to give a first look at a measurement process. After that, we give a mechanism for structure formation of a macroscopic system, also by forming an ``almost" quotient space, so that the macroscopic system's states behave more classically. A general description of a quantum measurement process is proposed in Sec.~\ref{secvi}, with the problem of the Schr\"odinger's cat also resolved by treating it as a particular measurement problem. A general comparison between our non-dynamical decoherence and the dynamical decoherence (or the einselection scheme) is given in Sec.~\ref{secvii}.  Finally, in Sec.~\ref{secviii}, we summarize our results and draw some conclusions and outlook.

\section{``Almost" Quotient Space and Non-dynamical Decoherence}
\label{secii}
\subsection{``Almost" quotient Hilbert space and equivalence classes}
\label{secii1}

For the total Hilbert space $\mathcal{H}_t$ in Eq.~(\ref{d}) that is roughly resolved into a studied or observed system and an environment, it is \emph{ergodic} from one particular state under all the possible dynamical evolutions. The number of those evolutions is large and can be collected as a set or a group $\{\hat{U}\}$ for short. Under this group, $\mathcal{H}_t$ is ``detected" in a \emph{fine grained} way.
Then, consider an observable $\hat{O}$ for the studied system in terms of orthonormal eigenstates
\begin{equation}
\hat{O}=\sum _{n}O_{n}\vert n\rangle\langle n\vert,\qquad \hat{O}\vert n\rangle=O_{n}\vert n\rangle.
\label{1}
\end{equation}
If performing a quantum measurement on a \emph{fixed or time independent} normalized state $\vert \psi\rangle$ on $\mathcal{H}_S$, we will obtain an average value
\begin{equation}
\bar{O}=\langle \psi\vert\hat{O}\vert \psi\rangle=\sum _{n}O_{n}|\alpha_{n}|^2,\qquad \vert \psi\rangle=\sum_{n}\alpha_n\vert n\rangle,
\label{2}
\end{equation}
with $|\alpha_{n}|^2$ treated as the probability for obtaining the state $\vert n\rangle$ after the measurement. If the operator $\hat{O}$ is an identity $\hat{I}=\sum _{n}\vert n\rangle\langle n\vert$, then we have
\begin{equation}
1=\langle \psi\vert\hat{I}\vert \psi\rangle=\langle \psi\vert \psi\rangle=\sum _{n}|\alpha_{n}|^2,
\label{3}
\end{equation}
which is the sum rule of probability. These are the standard results of quantum mechanics, named Born rule.

Now let's consider the problem in a different way. For Eq.~(\ref{3}), it is invariant under the group $\{\hat{U}\}$ of all the evolutions, since $[\hat{U},\hat{I}]=0$. This is also a standard result of quantum mechanics. However for Eq.~(\ref{2}), it is invariant only under some particular unitary evolutions, which can be collected as a smaller group
\begin{equation}
G_{\hat{O}}\equiv\left\{\hat{U},[\hat{U},\hat{O}]=0\right\}.
\label{4}
\end{equation}
$[\hat{U},\hat{O}]=0$ here serves as a stability condition for the observed system, which guarantees that the measurement can be accomplished without uncontrolled disturbances. In a quantum measurement, the macroscopic apparatus serves as an environment, thus we can first consider this particular case, with $\mathcal{H}_E$ in Eq.~(\ref{d}) replaced by the apparatus's Hilbert space $\mathcal{H}_A$. Then the evolutions in the group $G_{\hat{O}}$ stand for the interactions between the observed system and the large number of degrees of freedom of the apparatus. Certainly, there may be some perturbations that can relax the stability condition to $[\hat{U},\hat{O}]\simeq0$, but this does not change the main results. Hence, the group $G_{\hat{O}}$ should be treated as a necessary prerequisite to measure $\hat{O}$.
What is its effect on the observed system?

Let the group $G_{\hat{O}}$ operate on the space $\mathcal{H}_S\otimes\mathcal{H}^0_{A}$, in which the state is of some direct product form $\vert \psi\rangle\vert \phi\rangle^0_A$ with $\vert \psi\rangle$ and $\vert \phi\rangle^0_A$ the initial states of the system and apparatus respectively. Then a subspace $\mathcal{H}'$ of $\mathcal{H}_t$ can be obtained, since $G_{\hat{O}}$ is only a subgroup of the total one $\{\hat{U}\}$. There is a special property in $\mathcal{H}'$ that it is classified into some equivalence classes\footnote{The concept of equivalence class was also mentioned in~\cite{d4} for the problem of quantum gravity. } according to the group $G_{\hat{O}}$. For instance, let any two different evolutions $\hat{U}_1$ and $\hat{U}_2$ of the group $G_{\hat{O}}$ act on some initial state, then we have
\begin{equation}
\vert \psi\rangle\vert \phi\rangle^0_A\cong\hat{U}_1\vert \psi\rangle\vert \phi\rangle^0_A\cong\hat{U}_2\vert \psi\rangle\vert \phi\rangle^0_A.
\label{4a}
\end{equation}
This can be verified by substituting Eq.~(\ref{4a}) into Eq.~(\ref{2}), and note that $[\hat{U}_{1,2},\hat{O}]=0$. More precisely, we can express the space $\mathcal{H}'$ formally as
\begin{equation}
\mathcal{H}'\equiv G_{\hat{O}}(\mathcal{H}_S\otimes\mathcal{H}^0_{A})=\left\{\{G_{\hat{O}}\vert \psi\rangle\vert \phi\rangle^0_A\},\{G_{\hat{O}}\vert \varphi\rangle\vert \phi\rangle^0_A\},\cdots\right\},
\label{4b}
\end{equation}
provided that the expansion coefficients of the (fixed) states of $\mathcal{H}_S$ $\vert \psi\rangle=\sum_{n}\alpha_n\vert n\rangle$ and $\vert \varphi\rangle=\sum_{n}\beta_n\vert n\rangle$ satisfy
\begin{equation}
|\alpha_n|^2\neq|\beta_n|^2,\qquad for ~some ~n.
\label{4c}
\end{equation}
In a real measurement, a sample of the observed system's states should be prepared, one measurement for each prepared state. Then the initial state of the apparatus for each measurement should be ``almost" the same so that no significant disturbance occurs\footnote{According to decoherence, the prepared states for both the observed system and the apparatus should decohere from their respective environments. This is impossible in principle, since no system is isolated absolutely. However, in a practical sense, we can always assume that the prepared states are isolated initially, provided that the interactions from their environments are weak enough so that the correlations with the environments can be neglected.}. Here, we fix the apparatus's initial state for simplicity.

The complementary subspace $\mathcal{H}_t-\mathcal{H}'$ can be arrived at only by those evolutions not belonging to the group $G_{\hat{O}}$, i.e. by evolutions noncommutative with the observable $\hat{O}$. In this way, the total Hilbert space $\mathcal{H}_t$
\begin{equation}
\mathcal{H}_t=\mathcal{H}'\oplus(\mathcal{H}_t-\mathcal{H}')\simeq\mathcal{H}_t/G_{\hat{O}}
\label{5a}
\end{equation}
is ``detected" by those evolutions of the group $G_{\hat{O}}$ only in a \emph{coarse grained} way. The meaning of ``coarse grained" is as follows. The equivalence class $\{G_{\hat{O}}\vert \psi\rangle\vert \phi\rangle^0_A\}$ in Eq.~(\ref{4b}) can be treated as a ``state" in the space $\mathcal{H}'$, while the fine structure of the class is ``ignored" due to our inability to know the details of those evolutions in the group $G_{\hat{O}}$. Eq.~(\ref{5a}) is not an exact quotient space due to the complementary subspace $\mathcal{H}_t-\mathcal{H}'$, so we call it an ``almost" quotient space. The notation $\mathcal{H}_t/G_{\hat{O}}$ is used here only to denote the classification for short, similarly for some other $\mathcal{H}_t/G(.)$ in the following discussions, with $G(.)$ a possible subgroup of unitary evolutions. Since the environment here is only a measurement apparatus, thus we can concern only with $\mathcal{H}'$, otherwise no credible measurement could be performed.

\subsection{Non-dynamical decoherence: random phase approximation}
\label{secii2}

The representative state for one equivalence class in the space $\mathcal{H}'$ corresponding to some fixed state $\sum_{n}\alpha_n\vert n\rangle$ in $\mathcal{H}_S$ can be expressed as
\begin{equation}
\sum_{n}\alpha_n\vert n\rangle\vert \chi_n\rangle_{A}, \qquad \sum_n|\alpha_n|^2=1,
\label{6a}
\end{equation}
with $\{\vert \chi_n\rangle_{A}\}$ a collection of some normalized (\emph{not necessary orthogonal}) states for the apparatus. We can further make an expansion for each $\vert \chi_n\rangle_{A}$
\begin{equation}
\vert \chi_n\rangle_{A}=\sum_{i,j,k,\cdots}C_{n;i,j,k,\cdots}\vert i,j,k,\cdots\rangle_{A},
\label{7a}
\end{equation}
with $i,j,k,\cdots$ denoted as the large number of degrees of freedom of the apparatus. Here is just a general consideration, while a more detailed discussion with the apparatus's pointer states singled out will be shown in Sec.~\ref{secvi1}. There is a \emph{basis ambiguity} by transforming the set $\{\vert \chi_n\rangle_{A}\}$ into another set. As a consequence, the set of the observed system's states will also be transformed. For instance, by substituting Eq.~(\ref{7a}) into Eq.~(\ref{6a}), we have $\sum_{i,j,k,\cdots}(\sum_{n}\alpha_nC_{n;i,j,k,\cdots}\vert n\rangle)\vert i,j,k,\cdots\rangle_{A}$, where the linearly combination in the bracket is a new set of states for the observed system. This basis ambiguity is always present in principle for a \emph{closed} system's state, just like the case in Eq.~(\ref{a}). However, the ambiguity here is just a \emph{non-dynamical gauge freedom}, since if it had some dynamical sources, the involved evolutions would violate the stability condition $[\hat{U},\hat{O}]=0$ in the group $G_{\hat{O}}$ so that no credible outcomes could be obtained\footnote{For a representation transformation $\hat{S}$, the transformed observable $\hat{S}\hat{O}\hat{S}^{-1}$ is generally different from $\hat{O}$ unless $[\hat{S},\hat{O}]=0$. In this sense, the stability condition can also be used to choose a particular representation, removing the basis ambiguity.}. Thus, we can fix the gauge by choosing the expansion to be the form in Eq.~(\ref{6a}), removing the basis ambiguity.

One part of the quantum information is stored in the correlations of the system and the apparatus, in particular in the coefficients $\{C_{n;i,j,k,\cdots}\}$ of Eq.~(\ref{7a})
\begin{equation}
|C_{n;i,j,k,\cdots}|\exp(-i\gamma_{n;i,j,k,\cdots}).
\label{8a}
\end{equation}
The singled out phase factor gives the phase information exchanged between the observed system and the apparatus's degrees of freedom $i,j,k,\cdots$. In the large number(of the apparatus's degrees of freedom) limit, the phase in Eq.~(\ref{8a}) can be approximated to be random and depend only on the quantum number $n$
\begin{equation}
\alpha_ne^{-i\gamma_{n;i,j,k,\cdots}}\rightarrow\alpha_ne^{-i\gamma_{n}}\rightarrow|\alpha_n|e^{-i\gamma_{n}},
\label{9a}
\end{equation}
where the last step in Eq.~(\ref{9a}) indicates that the phase information stored in each $\alpha_{n}$ has been \emph{randomized}. This \emph{random phase approximation} concerns only with the average effects of the apparatus on the observed system, ignoring most of the details of the interactions. This is different from the Markovian approximation which ignores only the feedback of the system on the environment. Certainly, the reason for random phase approximation is due to our inability to keep track of the apparatus's large number of degrees of freedom. This implies that the random phase approximation serves as a further coarse grained approximation by ignoring some details of the information exchange. Moreover, this approximation depends little on the states $\{\vert \chi_n\rangle_{A}\}$ of the apparatus, while the dynamical decoherence approach needs a further orthogonality condition for those states. After the random phase approximation information is lost partially, just like the case in the eiselection scheme, inducing a decoherence.

This random phase approximation should be distinguished from the phase damping in dynamical decoherence, in which the relative phases in the non-diagonal terms of the density matrix are damped through the dynamical decoherence function in Eq.~(\ref{e}). \emph{The random phase approximation has no dynamical source, just serves as an approximative method to give a coarse grained description}, inducing a \emph{non-dynamical decoherence}.  Moreover, the decay term in Eq.~(\ref{e}) is mainly caused by some thermal or random properties of the environment, violating the Everettian picture. However, as we shall shown in the next subsection, the random phase approximation can still be described by some \emph{unitary} operators, preserving the Everettian picture. In a word, the random phase approximation or the randomness is introduced mainly because of our inability to keep track of the large number of the environment's degrees of freedom, not a necessary prerequisite of environment for inducing a decoherence.

It should still be emphasized that the studied information is actually not lost, but stored in those correlations implicit in Eq.~(\ref{6a}), in particular in those combined coefficients $\{\alpha_nC_{n;i,j,k,\cdots}\equiv D_{n;i,j,k,\cdots}\}$. Comparing a state $\sum_{n}\alpha_n\vert n\rangle$ in the space $\mathcal{H}_S$ with its corresponding equivalence class in the space $\mathcal{H}'$ we can see that, as an open system \emph{the information of the system's initial state stored in $\{\alpha_n\}$ has diffused into the correlations $\{D_{n;i,j,k,\cdots}\}$ with its surrounding environment, the apparatus here}. To recover the exact information of the system's initial state, we should know the precise details about the information exchange among those coupled systems. This seems to be impossible for any ordinary observer including our human beings, except for some kind of ``superobserver". For ordinary observers, the random phase approximation is thus introduced to obtain an effective coarse grained description.
We shall show in Sec.~\ref{secvi2} that this random approximation is also necessary for the macroscopic observations on the apparatus made by our human beings, since our undeveloped brains can not deal with the details of the information exchange\footnote{The brain is only treated as a physical structure composing of atoms, molecule and so on, which may also be regarded to be an information receiver. No meta-physical sense will be involved, as shown in Sec.~\ref{secvi2}.}.

The above discussions imply that, when making a quantum measurement, the large number of degrees of freedom of the apparatus leads to various evolutions in the group $G_{\hat{O}}$ and further results in a classification of the total Hilbert space, as given in Eq.~(\ref{5a}). This classification and a further random phase approximation provide a coarse grained description to the observed system. For another observable $\hat{O}'=\sum_{i}O'_{i}\vert i\rangle\langle i\vert$ that satisfies $[\hat{O}',\hat{O}]\neq0$, there will be another classification of the space $\mathcal{H}_t$ according to some group $G_{\hat{O'}}$. In this sense, the difference of measurement outcomes for two noncommutative observables comes from the fact that the relevant Hilbert space is classified in two different ways.

Now, let's consider a measurement on a time dependent state $\vert \psi(t)\rangle$ evolved via some evolution $\hat{U}(t)$ resulted from some other environment. In this case, an extra apparatus should be added to make a measurement. If $\hat{U}(t)$ is commutative with the observable $\hat{O}$, then it belongs to the group $G_{\hat{O}}$ and the above discussion is unchanged. If $\hat{U}(t)$ is noncommutative with $\hat{O}$, the Hilbert space will be some evolutionary one $\hat{U}(t)(\mathcal{H}^0_S\otimes\mathcal{H}^0_{E})$ from some particular initial space. At some instant, we can make a measurement by turning on the evolutions in the group $G_{\hat{O}}$, obtaining another $\mathcal{H}'$ formally expressed as $G_{\hat{O}}\hat{U}(t)(\mathcal{H}^0_S\otimes\mathcal{H}^0_{E}\otimes\mathcal{H}^0_{A})$. Then the mean value of the observable is given by
\begin{equation}
\bar{O}(t)=\langle \psi(0)\vert(\hat{U}\hat{U}(t))^{\dag}\hat{O}\hat{U}\hat{U}(t)\vert \psi(0)\rangle=\langle \psi(t)\vert\hat{O}\vert \psi(t)\rangle,
\label{10a}
\end{equation}
with some arbitrary $\hat{U}$ chosen from the group $G_{\hat{O}}$. This means that when making a (real) measurement, the evolution $\hat{U}(\tau)(\tau>t)$ should be stopped, otherwise no credible outcomes can be obtained. In other words, the operations of measurement and evolution cannot be turned on simultaneously. This is crucial for a quantum measurement, since the strengths of the measurement interactions or those evolutions in the group $G_{\hat{O}}$ are roughly equal to that of $\hat{U}(t)$\footnote{After one measurement, the evolution can be turned on again until next measurement. Continuing this step, we will obtain some statistical correlation function.}. While for a (classical) observation on a macroscopic system, for example our sense of vision, the observation is only a small perturbation to the evolution of the macroscopic system so that the operations of measurement and evolution may be turned on simultaneously.

One prepared state $\vert \psi(t_0)\rangle$ at some instant $t_0$ can be expanded as
\begin{equation}
\sum_{n}\alpha_n(t_0)\vert n\rangle\vert \chi_n(t_0)\rangle_{E}, \qquad \sum_n|\alpha_n(t_0)|^2=1,
\label{11a}
\end{equation}
with $\{\vert \chi_n(t_0)\rangle_{E}\}$ still a collection of some normalized (not necessary orthogonal) states for the environment. Analogous to Eqs.~(\ref{8a}) and~(\ref{9a}), the large number of degrees of freedom of the apparatus also provide an effective random phase to each $\alpha_n(t_0)$. In the next subsection, we will introduce a random phase unitary operator to describe this random phase approximation effectively, and derive the Born rule without using a dynamical decoherence function.

\subsection{Random Phase Unitary Operator and Born Rule}
\label{seciii}

The group $G_{\hat{O}}$ in Eq.~(\ref{4}) is not convenient to obtain measurement outcomes, because the group contains so many evolutions that it is difficult to be described. The random phase approximation in Eq.~(\ref{9a}) provides an effective description. Generally speaking, quantum information is exchanged under those evolutions in the group $G_{\hat{O}}$, and since phases are also important elements of quantum information, some phase factors can appear in those unitary evolutions of the group $G_{\hat{O}}$, just like the one in Eq.~(\ref{8a}). For a particular evolution, the phases of the observed system's states change regularly, but if there are a large number of evolutions, the phases will change in a random manner approximately. This usually occurs for a macroscopic system composing of large number of degrees of freedom, for example, a measurement apparatus, or some other general environment.
Thus we can introduce an effective random phase unitary operator acting only on the observed system's space
\begin{equation}
\hat{U}(\gamma)=\sum_{n}e^{-i\gamma_n}\vert n\rangle\langle n\vert,\qquad all ~\gamma_n ~is ~random,
\label{12a}
\end{equation}
with those random phases indicating an average information exchange for the observed system. Then there will be an limit
\begin{equation}
G_{\hat{O}}\longrightarrow G_{\gamma}\equiv\{\hat{U}(\gamma)\},
\label{13a}
\end{equation}
with $G_{\gamma}$ a group of all the random unitary operators with the form of Eq.~(\ref{12a}). The limit in Eq.~(\ref{13a}) is effective, as long as the number of the elements in the group $G_{\hat{O}}$ is large enough. Because $[\hat{O},\hat{U}(\gamma)]=0$, we can rewrite the observable $\hat{O}$ as
\begin{equation}
\hat{O}=\hat{U}^{-1}(\gamma)\hat{O}\hat{U}(\gamma).
\label{14a}
\end{equation}
This indicates that the observable operator $\hat{O}$ in fact does not act on a Hilbert space, but on a quotient space.

Suppose the Hilbert space of the system is
\begin{equation}
\mathcal{H}_S=\{\vert \psi\rangle,\vert \varphi\rangle,\cdots\},
\label{15a}
\end{equation}
then by means of the random phase unitary operator $\hat{U}(\gamma)$ or the group $G_{\gamma}$ in Eq.~(\ref{13a}), we will obtain an exact quotient Hilbert space
\begin{equation}
\mathcal{H}_S/G_{\gamma}=\left\{\{G_{\gamma}\vert \psi\rangle\},\{G_{\gamma}\vert \varphi\rangle\},\cdots\right\},
\label{16a}
\end{equation}
still as long as the expansion coefficients of the (fixed) states $\vert \psi\rangle=\sum_{n}\alpha_n\vert n\rangle$ and $\vert \varphi\rangle=\sum_{n}\beta_n\vert n\rangle$ satisfy
\begin{equation}
|\alpha_n|^2\neq|\beta_n|^2,\qquad for ~some ~n.
\label{17a}
\end{equation}
This gives an apparent coarse grained description of the original Hilbert space $\mathcal{H}_S$. Further, physical results do not depend on the random phases, thus some $\gamma$ invariant quantities should be constructed in the quotient space $\mathcal{H}_S/G_{\gamma}$. One simple example is the density matrix
\begin{equation}
\hat{\rho}\equiv\frac{1}{V(\gamma)}\int d\gamma\left[\hat{U}(\gamma)\vert \psi\rangle\langle \psi\vert\hat{U}^{-1}(\gamma)\right],
\label{18a}
\end{equation}
where $V(\gamma)$ is the volume of the $\gamma$ parameter space. After some simple calculations we will have
\begin{equation}
\hat{\rho}=\sum_{n}|\alpha_n|^2\vert n\rangle\langle n\vert,\qquad Tr\hat{\rho}=\sum_{n}|\alpha_n|^2=1,
\label{19a}
\end{equation}
which gives the Born rule for the probability. In deriving Eq.~(\ref{19a}), we have used
\begin{equation}
\int_{0}^{2\pi} d(\gamma_n-\gamma_m)e^{-i(\gamma_n-\gamma_m)}=0, (n\neq m),
\label{20a}
\end{equation}
because both $\gamma_n$ and $\gamma_m$ are random. Then by combining Eq.~(\ref{1}) and Eq.~(\ref{19a}), we can obtain the definite measurement outcomes
\begin{equation}
\{O_n,P_n=|\alpha_n|^2\}\rightarrow \bar{O}=\sum_{n}O_nP_n=Tr(\hat{O}\hat{\rho}).
\label{21a}
\end{equation}
This realizes our non-dynamical decoherence only through some random phase integrals, without involving any dynamical decoherence function of the form in Eq.~(\ref{e}).

It should be noted that the general space is actually $\mathcal{H}'$ in Eq.~(\ref{4b}), or the ``almost" quotient space in Eq.~(\ref{5a}) denoted by $\mathcal{H}_t/G_{\hat{O}}$. However, the space $\mathcal{H}_t/G_{\hat{O}}$ is difficult to be described due to the complexity of the group $G_{\hat{O}}$. After a further random phase approximation, a random phase unitary operator in Eq.~(\ref{12a}) together with the group $G_{\gamma}$ in Eq.~(\ref{13a}) is introduced, leading to an effective quotient space $\mathcal{H}_S/G_{\gamma}$ in Eq.~(\ref{16a}) that is easier to be described. In other words, the group $G_{\gamma}$ is a well parameterized approximation to $G_{\hat{O}}$ so that some quantities such as the density matrix in Eq.~(\ref{18a}) can be constructed effectively. These can be easily extended to the time dependent case in Eqs.~(\ref{10a}) and~(\ref{11a}), since a quantum measurement can be performed only on a state at some instant. Moreover, after the random phase approximation the phase information stored in each $\alpha_n$ is lost, as indicated by the randomization in Eq.~(\ref{9a}), while the probability information $|\alpha_n|^2$ is still retained and can be acquired by us, as will be shown in Sec.~\ref{secvi}.

There is a crucial difference between a Hilbert space and its corresponding (``almost") quotient space. In a Hilbert space, the superposition principle is applicable which roughly says that the space is closed under any linearly combinations of its states. However, in a (``almost") quotient Hilbert space, its elements are not simply states, but some equivalence classes with the form of Eq.~(\ref{16a}) or~(\ref{4b}). The question is then whether the superposition principle is also suitable for those classes, that is, \emph{whether any linearly combinations of the equivalence classes is still some single class}.
To understand this question easily, we take an example about the congruence concept in number theory, for instance $1\equiv8(mod~7)$, $3\equiv10(mod~7)$. That is, 1 and 8 belong to one equivalence class denoted as $\bar{1}$, while 3 and 10 belong to another class $\bar{3}$. Then we have $\bar{1}+\bar{3}=\bar{4}$, and this class addition is independent of the representative numbers.

Now consider the superposition of the equivalence classes in a (``almost") quotient space. For instance, we take the following three representative states from a single class $(e^{-i\alpha}\vert0\rangle+e^{-i\beta}\vert1\rangle)/\sqrt{2}$ in a quotient space of the form of Eq.~(\ref{16a})
\begin{equation}
\frac{1}{\sqrt{2}}(\vert0\rangle\pm\vert1\rangle),\frac{1}{\sqrt{2}}(\vert0\rangle+i\vert1\rangle).
\label{22a}
\end{equation}
Obviously, a direct addition of any two states will give states $\vert0\rangle$ or $\vert0\rangle+\frac{i\pm1}{2}\vert1\rangle$ (up to normalization constants), which belong to different classes. This indicates that the superposition principle is not suitable for equivalence classes. Certainly, for the classes of the eigenstates of some observable $\hat{O}$, any linear combination of them is indeed some single class. This is because the representative state for any eigenstate's class is just the eigenstate up to some irrelevant factors. However, this is just a special case, and the superposition can not be applied further for those general classes. Then it can be concluded that \emph{the superposition principle is not suitable for the elements or equivalence classes in a (``almost") quotient Hilbert space}. This can also be seen by noting that the stability condition in the group $G_{\hat{O}}$ in Eq.~(\ref{4}) breaks a lof of unitary symmetries. To recover the superposition principle, some evolutions noncommutative with the observable are needed to destroy the formed (``almost") quotient space.

Therefore, when making a measurement on a quantum system, some (``almost") quotient space can be formed according to some stability condition in  Eq.~(\ref{4}). Superposition principle, the most important quantum property will not be suitable for the equivalence classes in the (``almost") quotient space so that definite outcomes may be obtained. Moreover, in Sec.~\ref{secv} we shall show that, this unsuitable of superposition is crucial for macroscopic systems, since their states are also elements of some (``almost") quotient Hilbert space. After the measurement, the quantum property of the observed system may be recovered, provided that some other evolutions noncommutative with the observable $\hat{O}$ act on the system to destroy the formed ``almost" quotient space. This is not the case for a macroscopic system whose quantum properties may still be suppressed because of its stable structure, as will be shown in Sec.~\ref{secv2}.

\section{A two-dimensional example: Stern-Gerlach experiment}
\label{seciv}
As an example, we make a analysis on the Stern-Gerlach experiment, showing how the random phase unitary operator is introduced to obtain the measurement outcomes. Suppose the initial state of the electron's spin is $\vert \phi\rangle=\alpha\vert\uparrow_z\rangle+\beta\vert \downarrow_z\rangle$, and the interaction term is given by
\begin{equation}
\hat{H}_{int}=-\frac{e}{2m}\hat{\sigma}_{z}B_{z}(z),
\label{15}
\end{equation}
with $B_{z}(z)$ the z-component of a non-uniform magnetic field. With this interaction, we will have a state evolution
\begin{equation}
\vert \phi\rangle\stackrel{e^{-i\hat{H}_{int}t}}{\longrightarrow} \alpha e^{-i\omega t+ip_{z}z}\vert\uparrow_z\rangle+\beta e^{+i\omega t-ip_{z}z}\vert \downarrow_z\rangle,
\label{16}
\end{equation}
where $\omega =|e|B_{z}(0)/2m$, and
\begin{equation}
p_{z}z\simeq tFz \simeq-t\frac{|e|}{2m}\frac{dB_{z}}{dz}(0)z.
\label{17}
\end{equation}
The phase factors $e^{\pm ip_{z}z}$ indicate that there is momentum transfer between the electron and the magnetic field. Certainly, we can also assign $\vert\mp p_{z}\rangle_A$ as the momentum states of the apparatus or the magnetic field, and obtain the von Neumann's proposal~\cite{d}
\begin{equation}
\alpha e^{-i\omega t+ip_{z}z}\vert\uparrow_z\rangle\vert -p_{z}\rangle_A+\beta e^{+i\omega t-ip_{z}z}\vert \downarrow_z\rangle\vert +p_{z}\rangle_A,
\label{18}
\end{equation}
which gives the correlations between the electron and the measurement apparatus.

If the measurement apparatus is also a simple quantum system, the state in Eq.~(\ref{18}) is exact and no random phase will occur. However, this is not the case, because the magnetic field in Eq.~(\ref{15}) is only a classical quantity, and the full interaction should be in terms of some quantum filed operators
\begin{equation}
\hat{H}_{int}=-\frac{e}{2m}\hat{\sigma}_{z}\int d^3x\hat{\psi}^{\dag}(x) \hat{B}_{z}(x)\hat{\psi}(x).
\label{19}
\end{equation}
This means that when the electron is travelling in the magnetic field, it will always interact with the magnetic field by exchanging virtue photons so that its trajectory will seem to be random in a smaller scale. As a result, the momentum $p_{z}$ in Eq.~(\ref{17}) is only an average quantity, because $B_z=\langle\hat{B}_{z}\rangle$ has been used.
Then, a more precise phase factor will be
\begin{equation}
\exp(-i\omega t+ip_{z}z+i\gamma),
\label{20}
\end{equation}
with a random phase $\gamma$ resulting from the fluctuations due to the exchange of virtue photons.

The random phase $\gamma$ in Eq.~(\ref{20}) is not enough, since the phase factors for each term in Eq.~(\ref{16}) are not independent. This is because the evolution operator $e^{-iH_{int}t}$ in Eq.~(\ref{16}) belongs to $SU(2)$ satisfying $\det (e^{-iH_{int}t})=1$, since $Tr\hat{\sigma}_z=0$. For a general $U(2)$ group, there is not such a restriction. Now the states $\vert\mp p_{z}\rangle_A$ in Eq.~(\ref{18}) give a representation of the $U(2)$ group, so we can make an arbitrary $U(2)$ transformations $\hat{U}_A$ on them to obtain two new states. Since the states of the electron and the apparatus are entangled in Eq.~(\ref{18}), $\hat{U}_A$ will induce another $\hat{U}_{induce}$ transformation acting on the space of the electron. The transformation $\hat{U}_{induce}$ may be non-unitary because the electron's new states after the transformation may be non-orthogonal. This is the so called \emph{the problem of the preferred basis} or basis ambiguity in the quantum measurement, just like the case in Eq.~(\ref{a}). To obtain credible outcomes, we should remove this basis ambiguity coming from some general transformations involving linearly combinations of the states.

As shown below Eq.~(\ref{7a}), there is some non-dynamical gauge that should be fixed. After that, the possible general transformations $\hat{U}_A$ will be resulted from some dynamical evolutions. Recalling the group $G_{\hat{O}}$ defined in Eq.~(\ref{4}), the observable here is the electron's spin $\hat{S_z}$, so we should concern only with those evolutions satisfying $[\hat{U},\hat{S_z}]=0$. Thus $\hat{U}_A$ can not come from an interaction with the electron, since that would violate the condition $[\hat{U}_A,\hat{S_z}]=0$. The other source is from interactions with the rest degrees of freedom of the apparatus and a possible added environment. This means that the state in Eq.~(\ref{18}) should be extended to its corresponding equivalence class in an ``almost" quotient space like the one in Eq.~(\ref{5a}). For example, the representative state of the class corresponding to the first term in Eq.~(\ref{18}) is
\begin{equation}
\sum_{p_{z}}C_{\uparrow_z,p_{z}}\vert\uparrow_z\rangle\vert -p_{z}\rangle_A\vert\chi_{\uparrow_z,-p_{z}}\rangle_{res},
\label{21}
\end{equation}
with $C_{\uparrow_z,p_{z}}$ some transition amplitude resulted from the interactions between the electron and the apparatus, such as the phase factor in Eq.~(\ref{20}). Eq.~(\ref{21}) indicates that the states $\vert\mp p_{z}\rangle_A$ stand for only a small portion of the large number of degrees of freedom of the apparatus. Since the stability condition $[\hat{U}_A,\hat{S_z}]=0$ forbids the possible source from interaction with the electron, another condition $[\hat{U}_A,\hat{P}^A_z]=0$ is needed to forbid the possible interactions with the rest degrees of freedom. This is just the stability criterion~\cite{c} of the einselection scheme for selecting the observables of the apparatus by its environment. As will be shown in the next section, this stability condition is just one of a collection of conditions that can form a macroscopic system. Including the rest large number of degrees of freedom, a random phase approximation should be made according to the analysis below Eq.~(\ref{8a}). This can be simply achieved by using of a random phase unitary operator
\begin{equation}
\hat{U}_A=e^{-i\gamma_{1}}\vert -p_{z}\rangle\langle -p_{z}\vert+e^{-i\gamma_{2}}\vert +p_{z}\rangle\langle +p_{z}\vert,
\label{22}
\end{equation}
with the random phases $\gamma_1$ and $\gamma_2$ coming from the interactions with the large number of degrees of freedom of the apparatus and the possible environment. The above $\hat{U}_A$ can induce a random phase unitary transformation on the electron's spin space
\begin{equation}
\hat{U}_{induce}=e^{-i\gamma_1}\vert \uparrow_{z}\rangle\langle \uparrow_{z}\vert+e^{-i\gamma_{2} }\vert \downarrow_{z}\rangle\langle \downarrow_{z}\vert\equiv\hat{U}_{\gamma},
\label{23}
\end{equation}
just like the one of Eq.~(\ref{12a}). By means of this random phase unitary operator and Eq.~(\ref{18a}), we can obtain the required density matrix
\begin{equation}
|\alpha|^2\vert \uparrow_z\rangle\langle \uparrow_z\vert\vert-p_z\rangle_A\langle-p_z\vert+|\beta|^2\vert \downarrow_z\rangle\langle \downarrow_z\vert\vert+p_z\rangle_A\langle+p_z\vert,
\label{24}
\end{equation}
which contains only the classical correlations between the electron's spin states and the apparatus's pointer states, and gives the definite measurement outcomes.

\section{States of Macroscopic Systems}
\label{secv}
In common sense, a macroscopic system's states seem to be definite at some instant, i.e. its states behave classically. We can not perceive some superposition of states like $\alpha\vert alive\rangle+\beta\vert dead\rangle$ which says that a Schr\"odinger's cat is both alive and dead. The analysis in Secs.~\ref{secii} can be extended to describe the states of those macroscopic systems. Generally speaking, \emph{the states of a macroscopic system are not in a Hilbert space, but in a (``almost") quotient Hilbert space}.

\subsection{A toy model: classical bits from qubits}
\label{secv1}

Let's first consider a model for constructing classical bits from some qubits. Assume that there are N qubits and some interactions among them. Let the first qubit be a base with state $\vert \varphi\rangle=\alpha\vert0\rangle+\beta\vert1\rangle$, while the states for other qubits are all assuming to be $\vert0\rangle$ for simplicity. There is a class of primary and strong interactions (or evolutions) $\hat{U}_{1,i},i=2,\cdots,N$ which depends also on the distance of any two qubits. Then we can further assume a decreasing relation for the strengths of these interactions
\begin{equation}
\hat{U}_{1,2}\gg\hat{U}_{1,3}\gg\hat{U}_{1,4}\gg\cdots,
\label{25}
\end{equation}
that is, the qubits are ordered in increasing distances. There are also some much weaker interactions $\hat{U}_{i,j},i,j=2,\cdots,N$. This model is analogous to an atom with the first qubit as the atomic nucleus and the rest as electrons, and the couplings are the electromagnetic interactions. Consider first the first two qubits. With the interaction $\hat{U}_{1,2}$ turned on, we have a state evolution
\begin{equation}
\vert \varphi\rangle_1\vert 0\rangle_2\stackrel{\hat{U}_{1,2}}{\longrightarrow} \alpha e^{-i\gamma_{12}(t)}\vert0\rangle_1\vert0\rangle_2+\beta e^{-i\delta_{12}(t)}\vert1\rangle_1\vert1\rangle_2,
\label{26}
\end{equation}
i.e. a C-NOT gate~\cite{e} followed by a steady operator $e^{-i\gamma_{12}(t)}\vert 00\rangle_{12}\langle 00\vert+ e^{-i\delta_{12}(t)}\vert 11\rangle_{12}\langle 11\vert$, analogous to the steady states in an atom. Since the interaction $\hat{U}_{1,2}$ is the strongest, the structure in Eq.~(\ref{26}) should not be destroyed by the other interactions. In other words, the rest interactions should be perturbations satisfying (ideally) $[\hat{U}_{1,2},\hat{U}_{rest}]=0$. Then we can obtain a group of unitary interactions
\begin{equation}
G_{\hat{U}_{1,2}}\equiv\left\{\hat{U},[\hat{U}_{1,2},\hat{U}]=0\right\},
\label{27}
\end{equation}
which is analogous to the one in Eq.~(\ref{4}). With the stability condition in Eq.~(\ref{27}), the rest $\hat{U}_{1,i},i=3,\cdots,N$ can only be chosen to have the same form as $\hat{U}_{1,2}$, i.e. a C-NOT gate followed by a corresponding steady operator; while those $\hat{U}_{i,j},i,j=2,\cdots,N$ can only be some steady operators. Hence, after turning on the rest interactions, we will obtain a final state
\begin{equation} \alpha e^{-i\gamma}\vert0,0,\cdots,0\rangle_{1,\cdots,N}+\beta e^{-i\delta}\vert1,1,\cdots,1\rangle_{1,\cdots,N},
\label{28}
\end{equation}
with the random phase $\gamma$ (in the large N limit)
\begin{equation}
\gamma=\gamma_{12}+\gamma_{13}+\cdots\gamma_{1N}+\gamma_{i,j}+\gamma_{E}, (i,j=2,\cdots,N),
\label{29}
\end{equation}
similarly for the random phase $\delta$. The phase $\gamma_{E}$ is from the surrounding environment, since the couplings with the environment must also belong to the group $G_{\hat{U}_{1,2}}$ otherwise no stable structure would have been formed. Just like the case in Sec.~\ref{secii}, the state in Eq.~(\ref{28}) is just the representative state of an element in an effective quotient Hilbert space, an approximation of some ``almost" quotient space denoted as $\mathcal{H}_t/G_{\hat{U}_{1,2}}$. Here, $\mathcal{H}_t$ is the total Hilbert space of all the relevant coupled systems (N qubits) including the environment. Then by using of Eq.~(\ref{18a}), a statistic ensemble can be obtained
\begin{equation}
|\alpha|^2\vert \mathbf{0}\rangle\langle \mathbf{0}\vert+|\beta|^2\vert \mathbf{1}\rangle\langle \mathbf{1}\vert,
\label{30}
\end{equation}
with the classical bits
\begin{equation}
\vert0,0,\cdots,0\rangle_{1,\cdots,N}\equiv\vert\mathbf{0}\rangle,\vert1,1,\cdots,1\rangle_{1,\cdots,N}\equiv\vert\mathbf{1}\rangle.
\label{31}
\end{equation}

\subsection{Classical properties of macroscopic systems}
\label{secv2}

Although the above model is analogous to an atom, there are indeed some differences between them. In an atom, there are more energy levels than those of the qubits, and the condition in Eq.~(\ref{27}) may be relaxed to be some more physical one $G_{\hat{U}_{stru}}\equiv\left\{\hat{U},[\hat{U}_{stru},\hat{U}]\simeq0\right\}$
with $\hat{U}_{stru}$ responsible for the structure of an atom or even a macroscopic system. This means that some perturbations can deform the already formed structure a little. Besides, an atom can not contain too many electrons, then the phases are not random enough to obtain completely classical states. This is the reason for the quantum properties of an atom. However, atoms can be combined further to form molecules, macromolecules, crystals and even all the macroscopic systems. In these processes, the involved interactions are weaker and weaker so that the already formed atomic and molecular structures are preserved. Moreover, since the number of constituents is larger and larger, the information exchange among them is more and more complex, then the phases can be treated to be random so that the macroscopic systems' states behave more classically. In fact, a more general (ideal) group can be proposed as
\begin{equation}
G_{(\hat{U}^{a}_{stru},\hat{O}^{i}_{mac})}\equiv\left\{\hat{U},[\hat{U}^{a}_{stru},\hat{U}]=[\hat{O}^{i}_{mac},\hat{U}]=0\right\},
\label{31a}
\end{equation}
with $\{\hat{U}^{a}_{stru}\}$ the interactions responsible for all the structures of a macroscopic system, and $\{\hat{O}^{i}_{mac}\}$ standing for its \emph{macroscopic observables}, such as total mass, momentum, and other macroscopic properties. The indexes $a$ and $i$ are used to denote the collections for those operators, which satisfy the following commutative relations
\begin{equation}
[\hat{U}^{a}_{stru},\hat{U}^{b}_{stru}]=[\hat{O}^{i}_{mac},\hat{U}^{a}_{stru}]=[\hat{O}^{i}_{mac},\hat{O}^{j}_{stru}]=0.
\label{31b}
\end{equation}
This means that some quantum numbers can be assigned to denote the states of a macroscopic system.
With the help of Eq.~(\ref{31a}), the total Hilbert space can be classified into equivalence classes, obtaining an ``almost" quotient space in which the representative state for an equivalence class can be expressed as
\begin{equation}
\sum_{a,i,\mu_a,n_i}\alpha_{\mu_a,n_i}\vert \mu_a\rangle_{stru}\vert n_i\rangle_{mac}\vert \chi_{\mu_a,n_i}\rangle_{res},
\label{31c}
\end{equation}
with $\mu_a$ and $n_i$ the quantum numbers for the internal structures and the macroscopic observables of the macroscopic system respectively. Just like the states $\{\vert\chi_{n}\rangle_{A}\}$ in Eq.~(\ref{6a}), here $\{\vert \chi_{\mu_a,n_i}\rangle_{res}\}$ for the rest degrees of freedom are also some normalized states that can provide phase factors like the one in Eq.~(\ref{8a}). If the number of the rest degrees of freedom is large enough, a random phase approximation can also be made to give some statistical ensemble for the macroscopic system's states, leading to their classical properties.

The stability conditions in Eq.~(\ref{31a}) also provide a rough classification of all the possible interactions. The first class contains the weakest interactions satisfying all the conditions in Eq.~(\ref{31a}). For this class, the ``almost" quotient space is stable, that is, all the equivalence classes are invariant under those weakest interactions. The second class consists of some stronger interactions that may violate some condition for the macroscopic observables, but still satisfying those for the structures. Obviously, this class of interactions can only change the macro-states of the macroscopic system, such as momentum or angular momentum as a whole. The third class is the one consisting of the strongest interactions that violate most of the conditions in Eq.~(\ref{31a}), even those for the structures. Easily to see, the first two classes actually stand for the familiar ``classical" interactions, under which the ``almost" quotient space is stable or only deformed a little. While the third one can be described only by quantum theory, under which the ``almost" quotient space is no longer useful for describing the macroscopic system. Certainly, there may be some strongest interaction that violates only several conditions for the structures, for example, an X or gamma ray may destroy only some small structures of our larger bodies. In a word,\emph{ the ``almost" quotient space provides an effective border between the quantum and classical sides, both for the systems and the relevant interactions}. If there is no ``almost" quotient space in nature, the classical properties of macroscopic systems including our human beings, can not be emergent.

The above analysis implies that, the states of a macroscopic system are not simply elements of the total Hilbert space of all its constituents, but are elements of an ``almost" quotient space determined by some stability conditions in Eq.~(\ref{31a}). Besides, the structures of macroscopic systems could be formed only if the interactions were different in their strengths. This is the case in nowaday world, where there are four basic interactions whose strengths are indeed different and also depend on the distances. For instance, the strong interaction bounds quarks forming nucleons, while the weaker electromagnetic force bounds electrons and nucleons to form atoms. Since the electromagnetic force is weaker, so it cannot destroy the stable structures of the nucleons determined by the strong interaction. Further, as shown below Eq.~(\ref{22a}), the superposition principle is not suitable for the elements of a (``almost") quotient space, so it can be concluded that the states of macroscopic systems are more classical than quantum, as long as their stable structures are not destroyed by some other stronger interactions that violate most of the structure conditions in Eq.~(\ref{31a}). This is the case for those macroscopic systems in ordinary temperature, while in lower temperature some macroscopic quantum phenomena can occur, such as the superconductivity and superfluidity. This can be explained by noting that the classical properties are actually caused by the large number of the rest degrees of freedom in Eq.~(\ref{31c}). These degrees of freedom are \emph{independent} from each other in ordinary temperature, while in lower temperature they behave more and more \emph{dependently} so that their states' phases become coherent and can not provide random phases. Therefore, \emph{a macroscopic system behave classically if it contains large number of independent degrees of freedom}.

There is a general picture according to the previous discussions. Since no system is absolute, thus there is also no absolute Hilbert space for an observed system in the real case, except a total one $\mathcal{H}_t$ including all of the relevant coupled systems, perhaps the whole universe. If there is some local observable $\hat{O}$ or strongest interaction $\hat{U}_{stru}$ singled out, some ``almost" quotient space $\mathcal{H}_t/G_{\hat{O}}$ or $\mathcal{H}_t/G_{\hat{U}_{stru}}$ can be formed to describe a quantum measurement or a macroscopic system, as shown previously. The conditions in the groups $G_{\hat{O}}$ and $G_{\hat{U}_{stru}}$ imply that, the interactions violating those conditions has been neglected as long as they are weak enough not to disturb a measurement or a stable structure significantly. This is possible in the present age of our universe. Thus, \emph{a (``almost") quotient space is only phenomenal} in the sense that its formation depends greatly on the conditions of our universe. Besides, although an observed system or a macroscopic system described by some ``almost" quotient space behaves classically, the physical processes all behave in a quantum manner. Through a further random phase approximation by replacing the group $G_{\hat{O}}$ or $G_{\hat{U}_{stru}}$ with some corresponding random phase group $G_{\gamma}$, the ``almost" quotient space $\mathcal{H}_t/G_{\hat{O}}$ or $\mathcal{H}_t/G_{\hat{U}_{stru}}$ can be reduced to some effective quotient space $\mathcal{H}/G_{\gamma}$, with $\mathcal{H}$ the relevant Hilbert space for an observed system or a macroscopic system. The quotient space $\mathcal{H}/G_{\gamma}$ can thus be treated as the required space for the observed system or the macroscopic system, separated or decoherence from those large number of unobserved degrees of freedom effectively. Comparing with the abstract space $\mathcal{H}$, $\mathcal{H}/G_{\gamma}$ can only provide some statistical ensemble which gives definite measurement outcomes or some classical properties of a macroscopic system. Roughly speaking, \emph{although an abstract Hilbert space $\mathcal{H}$ is basic for describing a system, an open system can be described only by some (``almost") quotient Hilbert space}.

In an abstract Hilbert space, the superposition is applicable for its elements, but this is not the case for its corresponding (``almost") quotient space. Then, we can answer partly the question of the Schr\"odinger's cat. A superposition state of the form $\alpha\vert alive\rangle+\beta\vert dead\rangle$ for a Schr\"odinger's cat can exist only in an abstract Hilbert space, not in an ``almost" quotient space for a \emph{real} cat. Moreover, since the states $\vert alive\rangle$ and $\vert dead\rangle$ are the eigenstates of some macroscopic observable for that cat, a general state should be of the form in Eq.~(\ref{31c}), standing for one equivalence class in an ``almost" quotient space. A further analysis of the problem of the Schr\"odinger's cat will be given in the next section.

\section{Quantum Measurement Process}
\label{secvi}
A quantum measurement process is just an interaction between a quantum system and a macroscopic apparatus. As shown in Sec.~\ref{seciii}, the definite measurement outcomes are determined by some effective quotient Hilbert space in Eq.~(\ref{16a}). Moreover, we also show in Sec.~\ref{secv} that the states of a macroscopic system are elements of some ``almost" quotient Hilbert space, in which the superposition principle is not proper. It seems that the definite measurement outcomes are actually caused by the ``almost" quotient Hilbert space of the macroscopic apparatus. Here we give a general analysis.

\subsection{General descriptions of quantum measurement process}
\label{secvi1}

Suppose the observed system's Hilbert space is $\mathcal{H}_S$, while the ``almost" quotient Hilbert space of the apparatus is $\mathcal{H}_T/G_{\hat{U}_{stru},\hat{O}_{A}}$, with $\mathcal{H}_T$ the total space of the apparatus and a possible added environment. We can assume that the observed system and the apparatus are completely independent initially so that $\mathcal{H}_S$ is not in $\mathcal{H}_T$. $G_{\hat{U}_{stru},\hat{O}_{A}}$ is a group $\left\{\hat{U},[\hat{U},\hat{U}_{stru}]=[\hat{U},\hat{O}_{A}]=0\right\}$, where $\hat{U}_{stru}$ is the interaction responsible for the structure of the apparatus, and $\hat{O}_{A}$ is a macroscopic observable of the apparatus. Here we ignore the possible indexes of those operators for simplicity. Now, there is some interaction that can induce an evolution as
\begin{equation}
\mathcal{H}_S\otimes(\mathcal{H}_T/G_{\hat{U}_{stru},\hat{O}_{A}})\stackrel{\hat{U}_{int}}{\longrightarrow} (\mathcal{H}_S\otimes\mathcal{H}_T)/G_{\hat{O},\hat{U}_{stru}},
\label{32}
\end{equation}
giving a measurement on the observable $\hat{O}$ of the system. The group $G_{\hat{O},\hat{U}_{stru}}$, which is defined as $\left\{\hat{U},[\hat{U},\hat{O}]=[\hat{U},\hat{U}_{stru}]=0\right\}$, includes all the interactions that don't disturb the structure of the apparatus and the measurement outcomes, otherwise the measurement would be impossible. Eq.~(\ref{32}) implies that the ``almost" quotient space structure of the apparatus has been transferred into the combined system via the evolution $\hat{U}_{int}$ satisfying $[\hat{U}_{int},\hat{O}_A]\neq0$. It should be stressed that this evolution can not be a single basic unitary operator, because it also act on the apparatus's ``almost" quotient Hilbert space. Thus it must be a combination of a lot of basic evolutions constructed from the interactions between the observed system and the constituents of the apparatus and the possible environment
\begin{equation}
\hat{U}_{int}=\hat{U}_{1}\hat{U}_{2}\cdots\sim\exp(-i\sum_n\lambda_n\hat{P}_n\hat{T}_A),
\label{33}
\end{equation}
where $\hat{P}_{n}\equiv\vert n\rangle\langle n\vert$ is the projector corresponding to the observable $\hat{O}$ of the observed system. Moreover, $\hat{U}_{int}$ belongs to the group $G_{\hat{O},\hat{U}_{stru}}$, so $[\hat{T}_A,\hat{U}_{stru}]=0$. Thus $\hat{T}_A$ should be some \emph{macroscopic operator} acting on the apparatus's ``almost" quotient Hilbert space. It can only change the macro-states for the apparatus's macroscopic observable $\hat{O}_A$, inducing transitions on the ``almost" quotient space. For example, in the Stern-Gerlach experiment of Sec.~\ref{seciv}, $\hat{T}_A\simeq\hat{Z}_A$, the z-component of the apparatus's position (as a whole)~\cite{d}. It can lead to translations in momentum space obtaining the states $\vert\mp p_{z}\rangle_A$, the momenta of the apparatus as a whole.

The evolution in Eq.~(\ref{32}) can also be expressed as a transition of states
\begin{equation}
\vert n\rangle_S\vert O_A\rangle_{mac}\vert \chi_{O_A}\rangle_{res}\rightarrow\vert n\rangle_S\vert O^{(n)}_A(t)\rangle_{mac}\vert \chi_{O^{(n)}_A(t)}\rangle_{res},
\label{33a}
\end{equation}
where the apparatus's structure has been neglected for simplicity. The set of states $\{\vert O^{(n)}_A(t)\rangle_{mac}\}$ gives a representation of the unitary evolution $\hat{U}_{int}$. The time parameter $t$ indicates that those states are not exact eigenstates of the macroscopic observable $\hat{O}_A$, since $[\hat{O}_A,\hat{U}_{int}]\neq0$. After the measurement $\hat{U}_{int}$ is absent, then we will obtain another group of stability conditions $G_{\hat{O},\hat{U}_{stru},\hat{O}_{A}}$ and a corresponding ``almost" quotient space in which the representative state for an equivalence class can be expressed as
\begin{equation}
\sum_n\alpha_n\vert n\rangle_S\vert \phi_n\rangle_{T}=\sum_{n,O^{(n)}_A}\alpha_n\vert n\rangle_SC_{O^{(n)}_A}\vert O^{(n)}_A\rangle_{mac}\vert \chi_{O^{(n)}_A}\rangle_{res},
\label{33b}
\end{equation}
where the structure part is still neglected for simplicity. The state in Eq.~(\ref{33b}) has the same form as the one in Eq.~(\ref{6a}). Hence, we can make a further random phase approximation, by replacing the group $G_{\hat{O},\hat{U}_{stru},\hat{O}_{A}}$ with a corresponding (random phase) group $G_{\gamma}$ acting on either the system's space or a larger one including the macro-states $\{\vert O^{(n)}_A\rangle_{mac}\}$. By using of Eq.~(\ref{18a}), we will then obtain the required definite measurement outcomes. For the larger space, the coefficients $\{C_{O^{(n)}_A}\}$ can also contribute to the probabilities. These coefficients may be from the initial state of the apparatus, or from those transition amplitudes in Eq.~(\ref{33a}), i.e. $\langle n,O^{(n)}_A\vert\hat{U}_{int}\vert n,O_A\rangle$. In a real quantum measurement, the apparatus's initial state should be prepared to be almost the same for each measurement on the prepared system's states. Thus, the contribution from the apparatus's initial state can be removed by an initial random phase approximation, which prepares a definite state for the apparatus initially. While the contribution from the transitions can simply be treated as some measurement errors from the apparatus.

\subsection{Macroscopic observations}
\label{secvi2}

The expression in Eq.~(\ref{33b}) implies that the information of the system has been diffused into the correlations with the large number of degrees of freedom of the apparatus and the possible environment. Besides, the macro-states $\{\vert O^{(n)}_A\rangle_{mac}\}$ serve as the pointer states, pointing out the system's final state after the measurement. Then whether the full information of the system's initial state can be acquired by our human beings via some macroscopic observations on the apparatus?
Macroscopic observations are actually some interactions between two macroscopic systems. Those interactions should be weak enough to satisfy most of the stability conditions for both the two macroscopic systems, then we will obtain the ``classical" physical laws after some random phase approximations. For instance, our human beings, as a macroscopic system, can \emph{directly} observe another macroscopic system by only weak interactions, such as by \emph{the sense of vision, touch} and so on. The internal quantum properties of a macroscopic system can only be detected \emph{indirectly} via some quantum tool (for example, an X ray) that interacts strongly with the macroscopic system's constituents.

Further, the interactions between our human beings and a measurement apparatus should be so weak that even some of the apparatus's macro-states, for example the pointer states, can not be disturbed by our macroscopic observations, otherwise no credible outcomes can be obtained. As in Eq.~(\ref{33}), those weak interactions may be expressed as $\exp(-i\sum_{O_A}\lambda_{O_A}\hat{P}_{O_A}\hat{T}_{h})$ by combining a lot of basic interactions. $\hat{T}_{h}$ is a macroscopic operator acting on the ``almost" quotient space of our human beings, giving our pointer states just like the case of $\hat{T}_{A}$ for the apparatus. The final result is a complex correlation among the observed system, the measurement apparatus and our human beings. In particular, the system's eigenstates, the apparatus's pointer states and our pointer states are correlated in the same way as the expression in Eq.~(\ref{b}) for the einselection scheme. Therefore, our pointer states provide the perception about the state of the combined system-apparatus, obtaining the definite outcomes with the corresponding probabilities $\{|\alpha_nC_{O^{(n)}_A}|^2\}$. Then what about the remaining phase information stored in the phase factors of $\{\alpha_nC_{O^{(n)}_A}\}$?

Quantum information can be transferred between two simple quantum systems, for example two qubits. This can be achieved by a swap operation~\cite{e}
\begin{equation}
(\alpha\vert 0\rangle_a+\beta\vert 1\rangle_a)\vert 0\rangle_b\rightarrow\vert0\rangle_a(\alpha\vert 0\rangle_b+\beta\vert 1\rangle_b),
\label{33c}
\end{equation}
which also means that \emph{the qubit $b$ acquire the full information of the qubit $a$}. However, if the qubit $b$ is replaced by a macroscopic system composing of a lot of independent degrees of freedom, there will exist various interactions other than the simple swap operation, disturbing the information transfer significantly. The result is that \emph{the information stored by one degree of freedom will be shared by almost all the independent degrees of freedom of the macroscopic system}. This is analogous for a quantum measurement, as indicated by the state in Eq.~(\ref{33b}). A definite measurement outcome contains the full quantum information, only if the system's initial state was just an eigenstate of the observable. If not, correlations must be established between the system and the apparatus, where the quantum information is stored, i.e. the information of the system's initial state has been diffused into the apparatus. This is a general result. In order to acquire the full quantum information of the system, a \emph{recovery operation} should be performed. If the apparatus was also a simple quantum system, this recover operation may be constructed easily, one simple choice is the inverse of the evolution operator $\hat{U}_{int}$. For example, the entangled state $\alpha\vert 0\rangle_{S}\vert 0\rangle_{A}+\beta\vert 1\rangle_{S}\vert 1\rangle_{A}$ can be obtained from an initial one $(\alpha\vert 0\rangle_{S}+\beta\vert 1\rangle_{S})\vert 0\rangle_{A}$ via a C-NOT gate~\cite{e}. Then by performing the inverse of the C-NOT gate, the information of the system can be recovered. This may also be treated as an evidence for the symmetry under $t\rightarrow-t$ or the \emph{reversibility} of any basic unitary evolution.

However, for the complex evolution in Eq.~(\ref{33}) that is a combination of a lot of basic evolutions, its corresponding recovery operation is too difficult to obtain, because the details of those basic evolutions are very complicated. Thus, to construct such a recovery operation, a \emph{superobserver} is needed to keep track of the full details among those basic evolutions. This can also be seen from the view of the (``almost") quotient space. Note that the elements in an ``almost" quotient Hilbert space are some equivalence classes for which the superposition principle is not suitable. Thus, transitions among those classes can not be induced by a single basic unitary evolution, but only by a combination of some primary evolution together with a lot of evolutions belonging to some group $G(.)$, with ``$.$" denoted as some stability conditions. This means that the group of all unitary evolutions $\{\hat{U}\}$ can also be classified into some equivalence classes or cosets, obtaining a quotient space $\{\hat{U}\}/G(.)$. This is also an abstract description for the rough classification of all the interactions given in the last section. In this sense, the evolution in Eq.~(\ref{33}) actually stands for some equivalence class or coset of
a quotient space $\{\hat{U}\}/G_{\hat{U}_{stru},\hat{O}_{A}}$. In one word, the complexity of those interactions make the symmetry under $t\rightarrow-t$ broken \emph{phenomenally}, leading to the \emph{irreversibility} for macroscopic systems. This description of appearance of time direction may also be applied in statistical mechanics, leading to the second law of thermodynamics. Analogously, a recovery operation is also impossible \emph{phenomenally}, that is, it can not be constructed in a practical sense.

Therefore, a macroscopic observation made by us can not be a recovery operation to acquire the full information of the system, especially the phase information. One may ask whether we can observe a quantum system directly or via some quantum tool instead of a macroscopic one. Analogous to the previous discussions about the quantum measurement process, a direct observation or observation via a quantum tool can indeed establish the required correlations. That is, the information of the system or the system-tool can also be diffused into our bodies, since our human beings are also macroscopic systems. However, \emph{the brains of our human beings are not developed enough to deal with the quantum information stored in those correlations, thus a random phase approximation is necessary for us to obtain only a coarse grained description}. This can also be seen as follows. Although all the physical processes behave in a quantum manner, our brain as a macroscopic system is always described by an ``almost" quotient space so that all the phase information is still hidden. To acquire the phase information, the brain's ``almost" quotient space should be destroyed by some stronger interaction. Then our brain will also be damaged so that nothing can be perceived by us. In this sense, the random phase approximation is inevitable for us to perceive the physical world. As a consequence, a feeling that the information of the observed system is lost in a quantum measurement is only \emph{phenomenal}. All the information is still stored in those correlations among the observed system, the measurement apparatus(including a possible added environment) and our human beings. \emph{It is our undeveloped brain or its ``almost" quotient space structure that prevents the full quantum information from being acquired by us}. The quantum measurement problem is thus resolved, similarly for those classical properties of our surroundings.

According to the above analysis, the problem of the Schr\"odinger's cat can also be resolved in an analogous way. As a macroscopic system, the cat's states are also elements of some ``almost" quotient Hilbert space, in particular, the states $\vert alive\rangle$ and $\vert dead\rangle$ are the eigenstates of some macroscopic observable for that cat. The cat is coupled with a two level atom($E_2>E_1$) via a complex evolution or coupling $\hat{U}_{int}\sim e^{-i\sum_E\lambda_E\hat{P}_E\hat{T}_{cat}}$ like the one in Eq.~(\ref{33}). Following Eq.~(\ref{32}) or Eq.~(\ref{33a}), we then have an evolution or a state transition
\begin{equation}
\vert E_2\rangle\vert alive\rangle\stackrel{\hat{U}_{int}}{\longrightarrow}e^{-i\gamma}\alpha\vert E_2\rangle\vert alive\rangle+e^{-i\delta}\beta\vert E_1\rangle\vert dead\rangle,
\label{34}
\end{equation}
with $\alpha$ and $\beta$ some transition amplitudes for the atom's states. The parts for the structures of the cat and its rest degrees of freedom have been ignored for simplicity. The random phase factors $e^{-i\gamma}$ and $e^{-i\delta}$ are provided by the large number of those rest degrees of freedom of the cat and a possible added environment via a random phase approximation\footnote{Similar description of Schr\"odinger's cat and the random phase approximation was also given in~\cite{e1}, where some kind of ontological states was used.}. Hence, we will obtain a statistical ensemble with only classical correlations between the atom and the cat. In this sense, the cat can just be treated as a measurement apparatus to detect the states of the atom.

In conclusion, \emph{a quantum measurement is just a complex ``unitary" interaction or evolution} constructed by combining a lot of basic unitary evolutions, as given by Eq.~(\ref{33}). Different from those basic evolutions, this complex evolution also acts on the ``almost" quotient space of the macroscopic apparatus, i.e. it is actually an element of some quotient space $\{\hat{U}\}/G(.)$ for the unitary evolutions. Besides, this interaction establishes a correlation between the system's measured observable $\hat{O}$ and a macroscopic one $\hat{O}_A$ of the apparatus. When the apparatus is observed by our human beings, the involved interactions lead to a more complex correlation among the system's eigenstates, the pointer states of the apparatus and our human beings. A random phase approximation is needed because the interactions involved in the macroscopic observation can not recover the full quantum information, especially the phase information. As a result, only definite outcomes with the corresponding probabilities can be acquired by us. This means that \emph{there is an intrinsical limit for our human beings not to acquire the full (phase)quantum information, due to our brain's stable ``almost" quotient space structure}. This agrees with the reason for introducing decoherence, the inability to keep track of the large number of (independent) degrees of freedom of the relevant environment.

\section{Dynamical or Non-dynamical: Comparisons with Einselection}
\label{secvii}
It's worth to make a comparison between our non-dynamical decoherence or random phase approximation and the dynamical decoherence or the einselection scheme. The same point is that both of them use the concept of decoherence, i.e. the observed system is open to its surroundings. But the methods to induce a decoherence are different, dynamical decay in Eq.~(\ref{e}) for the einselection scheme while a random phase approximation for our approach. The condition for the dynamical decay in Eq.~(\ref{e}) seems to be more strict, indicated by the orthogonality condition $\langle E_1\vert E_2\rangle\simeq0$ for environment's states. This makes the dynamical decoherence or einselection scheme concern mainly with some ``local" details of the coupled systems, such as the possible forms of the relevant interactions and states, by proposing some \emph{particular} models to describe the decoherence~\cite{a,c,d1,f,g,h}. While our approach provides a ``global" view about some general properties of an \emph{arbitrary} open system, by forming an ``almost" quotient Hilbert space according to some stability conditions. When forming that ``almost" quotient space, a lot of unitary symmetries are broken so that the superposition principle is no longer proper generally. As a result, definite measurement outcomes may be obtained via a further random phase approximation.
A detailed comparison is as follows:

(i) In the einselection scheme, when deriving the Born rule Eq.~(\ref{c}) a partial trace is used. This seems to be a \emph{circular argument}~\cite{a,b} since the partial trace itself can also be treated as one part of the Born rule. That is, a partial trace over the environment's space can be regarded to be another measurement on the environment. For this reason, Zurek introduced the concept of envariance~\cite{a,c,i,j}, i.e. environment-assisted invariance to give another derivation of the Born rule, but his method is not general enough to be used wildly. In our approach, an effective quotient space in Eq.~(\ref{16a}) is obtained after a random phase approximation. In that space, an integration over the random phases' space is performed to obtain the Born rule, as given by Eq.~(\ref{18a}). This does not involve any circular argument.

(ii) In the einselection scheme, the apparatus's pointer states are \emph{selected} or \emph{determined} by the form of the interaction between the apparatus and its environment. It seems that the interaction with the environment plays a \emph{dominate} role in the measurement process. However, from Eq.~(\ref{a}) to Eq.~(\ref{c}) there is an implicit \emph{order} that the interaction between the system and the apparatus is first turned on. Although in the real case the two interactions actually occur independently, the implicit order indicates that \emph{the interaction from the environment is only a perturbation}. In our approach, the apparatus's pointer states are \emph{determined} mainly by the interaction between the system and the apparatus, but are \emph{stabilized} by some more stability conditions that may involve a possible added environment, as shown in Sec.~\ref{secvi1}.

(iii) In the einselection scheme, the orthogonality condition $\langle E_1\vert E_2\rangle\simeq0$ or the decoherence function in Eq.~(\ref{e}) for the environment's states are not satisfied generally, though in some controlled model this condition may be achieved approximately by assuming the thermal or random properties of the environment. In a more general case, the environment is uncontrolled so that the orthogonality condition can not be proposed without any question. In our approach, the random phase approximation is effective as long as the number of the relevant environment's (independent) degrees of freedom is large enough. When the environment is also a simple quantum system or lives in lower temperature, random phase approximation is no longer proper. In this case, no decoherence can occur except that an extra observation is made by us, involving a macroscopic apparatus. Therefore, in our approach, decoherence is only an approximative description, instead of a dynamical process as in the einselection scheme or dynamical decoherence, see the discussions below Eq.~(\ref{9a}).

(iv) The einselection scheme doesn't provide an effective description of macroscopic systems. For example, for a macroscopic apparatus in a quantum measurement, all the scheme says is that the apparatus decoheres in some basis due to interactions with its environment. Even the dynamical decoherence function in Eq.~(\ref{e}) seems to depend on some prerequisite classical properties of the macroscopic environment, violating the Everettian picture~\cite{d3}. In fact, to the observed system, the apparatus also serves as an environment because of its large number of degrees of freedom. The pointer states stand for only a very small portion of the apparatus, with the rest degrees of freedom unobserved. In our approach, these unobserved degrees of freedom help to form the (``almost") quotient space of the macroscopic apparatus, and also provide approximative random phases to lead to the apparatus's classical properties, as shown in Sec.~\ref{secv}.

\section{Conclusions and Outlook}
\label{secviii}

In this paper, we have shown that a quantum measurement is a complex ``unitary" interaction, with the complexities coming from the large number of independent degrees of freedom of the macroscopic apparatus and a possible added environment. The classical properties of the macroscopic apparatus are mainly caused by its large number of (unobserved) degrees of freedom, and can be described by the emergence of an ``almost" quotient Hilbert space phenomenally, determined by some necessary stability conditions. After the measurement, the combined system-apparatus can still be described by another ``almost" quotient Hilbert space with some classical properties. Macroscopic observations made by our human beings can not recover the full quantum information, especially the phase information. A random phase approximation is thus needed to give only a coarse grained description, inducing a non-dynamical decoherence. As a result, only the information stored in the remaining classical correlations among the observed system, the apparatus and our human beings, can be acquired by us. This gives the definite measurement outcomes with the corresponding probabilities. Quantum measurement problem is resolved.

As shown in this paper, the concepts of ``almost" quotient Hilbert space and equivalence class are useful to describe the classical properties of macroscopic systems in a general way. Whether they can be used to describe some particular systems in some more detailed way will still need to be investigated. For example, in describing the gravity, whether those concepts can be applied to explain why the gravitational field in our surrounding behaves more classically, or why the gravity is so difficult to be quantized? According to the previous analysis, there should be some stability condition so that the gravitational field can decohere from the rest still quantized matter fields and obtain some classical properties. What is that required stability condition is not known to us.

\section*{Acknowledgments}
This work is supported by the NNSF of China, Grant No. 11375150.


\end{document}